\documentclass[10pt]{iopart}

\expandafter\let\csname equation*\endcsname\relax
\expandafter\let\csname endequation*\endcsname\relax
\usepackage{iopams}  

\usepackage{graphicx}
\usepackage{dcolumn}
\usepackage{xcolor}
\usepackage{float}
\usepackage{bm}
\usepackage{comment}
\usepackage{tabularx}
\usepackage{cite}
\usepackage{microtype}

\usepackage[caption=false]{subfig}
\usepackage[italicdiff]{physics}
\usepackage[colorlinks=True,allcolors=blue,hyperfootnotes=false]{hyperref}

\allowdisplaybreaks
\begin{document}

\title{Correlations and signaling in the Schr\"{o}dinger--Newton model}

\author{Jacek Aleksander Gruca$^{1,2}$\footnotemark[1],
Ankit Kumar$^{3,4,1}$\footnotemark[1],
Ray Ganardi$^{1,5,6}$,
Paramasivan Arumugam$^{3,7}$,
Karolina Kropielnicka$^{8}$, and
Tomasz Paterek$^{1,9}$}

\footnotetext[1]{These authors contributed equally to this work.}

\address{$^{1}$ Institute of Theoretical Physics and Astrophysics, Faculty of Mathematics, Physics and Informatics, University of Gda{\'n}sk, 80-308 Gda{\'n}sk, Poland}

\address{$^{2}$ Department of Applied Physics and Applied Mathematics,  Columbia University in the City of New York, New York 10027, United States of America}

\address{$^{3}$ Department of Physics, Indian Institute of Technology Roorkee, Roorkee 247667, India}

\address{$^{4}$ International Centre for Theory of Quantum Technologies, University of Gda{\'n}sk, 80-309 Gda{\'n}sk, Poland}

\address{$^{5}$ Centre for Quantum Optical Technologies, Centre of New Technologies, University of Warsaw, 02-097 Warsaw, Poland}

\address{$^{6}$ School of Physical \& Mathematical Sciences, Nanyang Technological University, Singapore 637371, Singapore}

\address{$^{7}$ Centre for Photonics and Quantum Communication Technology, Indian Institute of Technology Roorkee, Roorkee 247667, India}

\address{$^{8}$ Institute of Mathematics, Polish Academy of Sciences,
81-825 Sopot, Poland}

\address{$^{9}$ School of Mathematics and Physics, Xiamen University Malaysia, 43900 Sepang, Malaysia}

\ead{kumar.ankit.vyas@gmail.com}

\begin{abstract}
The Schr\"odinger--Newton model is a semi-classical theory in which, in addition to mutual attraction, massive quantum particles interact with their own gravitational fields. While there are many studies on the phenomenology of single particles, correlation dynamics in multipartite systems is largely unexplored. Here, we show that the Schr\"odinger--Newton interactions preserve the product form of the initial state of a many-body system, yet on average agreeing with classical mechanics of continuous mass distributions. This leads to a simple test of the model, based on verifying bipartite gravitational evolution towards non-product states. We show using standard quantum mechanics that, with currently accessible single-particle parameters, two masses released from harmonic traps get correlated well before any observable entanglement is accumulated. Therefore, the Schr\"odinger--Newton model can be tested with setups aimed at observation of gravitational entanglement with significantly relaxed requirements on coherence time. We also present a mixed-state extension of the model that avoids superluminal signaling.
\end{abstract}

\maketitle

\section{Introduction}
Scenarios where quantum mechanics and gravity simultaneously play a role are not yet well understood~\cite{Aspelmeyer2022,Carlesso2022,bose2023massive,Donadi2022,Christodoulou2019}.
The Schr\"odinger--Newton (SN) model is a particular way of treating non-relativistic gravitating quantum systems that offers a departure from canonical quantum predictions in the regime of mass larger than $10^{-18}$ kg~\cite{Giulini_2011}.
The model can be derived in linearized gravity with the semi-classical assumption of replacing the stress-energy tensor with a particular expectation value~\cite{Bahrami2014}.
Most of the theoretical work focused on analyzing the single-particle SN dynamics~\cite{moller1962theories,Rosenfeld1963On,Disi1984,Penrose1996,Moroz1998,Tod1999,Adler2007,Carlip2008,Giulini2011,Colin2016,Groardt2016approx,Kelvin2020,giulini2011gravitationally,Guzmn2004}, 
and showed that the experiments testing it require a relatively large mass and long coherent evolutions~\cite{ReviewBassi}.
These are contradictory demands that are yet to be resolved in laboratories.

In this work we show that a bipartite setup provides a more straightforward and accessible test of the SN model: 
the SN equation preserves the product form of the initial state while recovering canonical gravity on average.
Consequently, an experimental confirmation of gravitational evolution into non-product states would disqualify the model.
These non-product states could be mixed and hence not necessarily entangled, simplifying experimental arrangements.
We compute the range of parameters for two masses released from harmonic traps, similar to those recently cooled in Vienna~\cite{Aspelmeyer2020-Science}, to argue that such a test is feasible.
Other optomechanical tests of the SN model have been proposed~\cite{Groardt2016,Groardt2016harmonic}, including in the limit of many particles~\cite{Yang2013Macroscopic}. 
They differ from this proposal as they are either based on single-particle properties or features of the center-of-mass of macroscopic objects.

The SN model is a nonlinear and deterministic modification of the Schr\"odinger equation.
It was shown by Gisin that any
deterministic and nonlinear theory
allows for superluminal signaling~\cite{Gisin_1989}.
This is a strong theoretical argument against this model, but it can be avoided by upgrading the dynamical equation to a stochastic one~\cite{Nimmrichter2015}.
We show towards the end of the paper, that the signaling problem can be avoided without stochastic calculus by extending the evolution to ensembles of pure states.

\section{Bipartite SN Dynamics}
\label{section:BipartiteDynamics}

Let us introduce the model gradually. For a single free massive point particle, described by wave function $\psi(\vec{r},t)$, the SN equation contains a nonlinear term that describes the self-gravitation:
\begin{equation}
i\hbar\frac{\partial}{\partial t} \psi(\vec{r},t) = \qty( -\frac{\hbar^2}{2m}\nabla^2 + V_\text{SN}(\vec{r},t) ) \psi(\vec{r},t),
\label{EQ_1SN}
\end{equation}
where the nonlinear part can be interpreted as a Newtonian coupling of the point mass $m$ with mass distribution generated by its own probability density $m|\psi|^2$:
\begin{equation}
    V_\text{SN}(\vec{r},t) = - Gm \int d \vec{r}\,' \ \frac{m|\psi(\vec{r}\,',t)|^2}{|\vec{r}-\vec{r}\,'|} .
\end{equation}

\begin{figure}
\centering
\includegraphics[width=\linewidth]{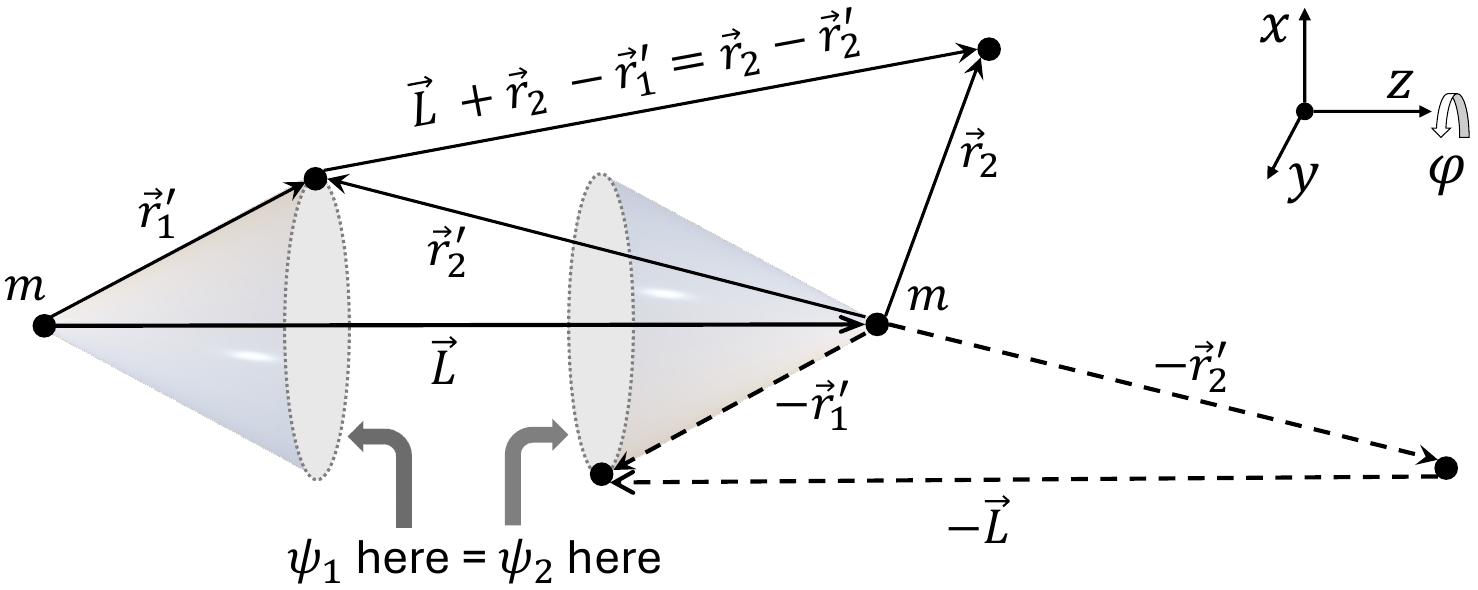}
\caption{Symmetry of the problem. Two identical masses $m$ are initially placed a distance $L$ apart. The system has axial symmetry, and therefore, the wave function of the left mass, called $\psi_1$, is equal to the wave function of the right mass, $\psi_2$, at the indicated cones. The explicit variables are used to decouple Eqs.~(\ref{EQ_COUPLED}). Vectors $\vec r_j$ and $\vec{r}_j '$ are counted from the initial average position of the $j$th particle.}
\label{FIG_GEO}
\end{figure}

Consider now two point particles separated by a distance $L$ along the $z$ axis.
For simplicity, we assume both particles have the same mass $m$, and are initially prepared independently in pure states: $\Psi(\vec r_1, \vec r_2, 0) = \psi_1(\vec r_1, 0) \, \psi_2(\vec r_2, 0)$.
We show that within the SN model, the product form of the initial state is preserved at all times.
In the bipartite setting each particle interacts with its own gravitational field and with the field of the other particle. The potential energy is, therefore, a sum of four terms~\cite{Disi1984,Adler2007}:
\begin{equation}
V(\vec{  r}_1,\vec{  r}_2, t) = V_{11} + V_{12} + V_{21} + V_{22},
\label{EQ_V}
\end{equation}
where the self-interaction terms are
\begin{equation}
V_{jj} = - G m \int d \vec{r}_1\!' d \vec{r}_2\!' \ \frac{m|\Psi(\vec{r}_1\!', \vec{r}_2\!', t)|^2}{ |\vec{r}_j-\vec{r}_j\!'|} ,
\label{pot-self}
\end{equation}
and the mutual interaction terms are given by
\begin{align}  
V_{12} =& -Gm\int d\vec{r}_1\!'d\vec{r}_2\!'\frac{m|\Psi(\vec{r}_1\!',\vec{r}_2\!', t)|^{2}}{|\vec{r}_1-\vec{r}_2\!'-\vec{L}|}  ,
\nonumber\\[0.5em]
V_{21} =& -Gm\int d\vec{r}_1\!'d\vec{r}_2\!'\frac{m|\Psi(\vec{r}_1\!',\vec{r}_2\!', t)|^{2}}{|\vec{r}_2-\vec{r}_1\!'+\vec{L}|} ,
\label{pot-mutual}
\end{align}
where $\vec{L}$ points from the initial average position of particle $1$ to the initial average position of particle $2$
(see Fig.~\ref{FIG_GEO}).
Note that in every potential energy term $V_{jk}$ the primed variables are integrated out, and thus every term is a function of either $\vec r_1$ or $\vec r_2$, but never of both.
Accordingly, the full SN Hamiltonian is the sum $H_1(t) + H_2(t)$, where each $H_j(t)$ is nonlinear and implicitly time-dependent, but involves only the variables of the $j$th particle. Consequently, $H_1$ and $H_2$ commute at all times.
Assuming the masses are prepared in a Gaussian state, in~\ref{appendix:SN-Pot-InitialTime}  we calculate the explicit form of these interaction terms at $t=0$. 
This separability of the Hamiltonian into the sum of distinct commuting parts implies that the product form of the initial state is maintained at all times:
\begin{equation}
\psi_1(\vec r_1,0) \, \psi_2(\vec r_2,0) \to \psi_1(\vec r_1, t) \, \psi_2(\vec r_2, t).
\end{equation}
We prove this carefully in~\ref{appendix:SN-ProductForm-Proof}.

The preservation of the product form and axial symmetry of the problem can be exploited to simplify the bipartite SN equation into an effective single-particle equation, which will be used to prove the classical limit.
For brevity, let us denote $\psi_j = \psi_j(\vec r_j,t)$.
The SN equation for the product state reads:
\begin{equation}
i \hbar \pdv{t} \qty( \psi_1 \, \psi_2 )  =   
- \frac{\hbar^2}{2 m} \Big( (\nabla_1^2 \psi_1) \, \psi_2  + \psi_1 \, ( \nabla_2^2 \psi_2) \Big)
\\
+ (V_{11} \psi_1 + V_{12} \psi_1) \psi_2 
+ \psi_1 (V_{21} \psi_2 + V_{22} \psi_2).
\label{eq:SNEqFull}    
\end{equation}
One can now use the product rule on the left-hand side and divide by $\psi_1 \psi_2$ to see that the separation of variables leads to:
\begin{align}
i \hbar \pdv{t} \psi_1  =& 
\left( - \frac{\hbar^2}{2 m} \nabla_1^2
+ V_{11} \qty[|\psi_1|^2] + V_{12} \qty[|\psi_2|^2] + C \right) \psi_1,
\nonumber \\[0.5em]
i \hbar \pdv{t} \psi_2  =& 
\left(  - \frac{\hbar^2}{2 m} \nabla_2^2 + V_{21}\qty[|\psi_1|^2] + V_{22}\qty[|\psi_2|^2] - C
\right) \psi_2,
\label{EQ_COUPLED}
\end{align}
where $C$ is independent of $\vec r_1$ and $\vec r_2$, and for simplicity we put $C = 0$.
These equations are coupled and the square brackets indicate the functions on which potential energies $V_{jk}$ depend on. 
For example, $V_{21}\qty[|\psi_1|^2]$ depends on the state of the first particle:
\begin{equation}
V_{21}
= - G m \int d \vec{r}_1\!' \ \frac{m|\psi_1(\vec{r}_1\!', t)|^2}{|\vec{r}_2 - \vec{r}_1\!'+\vec{L}|},
\label{EQ_V21}
\end{equation}
where we used normalization $
\int d\vec{r}_2\!' \abs{\psi_2(\vec{r}_2\!',t)}^2 = 1$.
However, since the bipartite system admits axial symmetry, the wave functions $\psi_1$ and $\psi_2$ are independent of the azimuthal angle $\varphi$ and, assuming they are initially the same, they are mirror reflections about the $xy$ plane that bisects the symmetry axis. 
Fig.~\ref{FIG_GEO} shows the relevant variables and properties that justify the following steps:
\begin{eqnarray}
V_{21}
& = & - G m \int d \vec{r}_2\!' \ \frac{m|\psi_1( \vec{r}_2\!'+\vec{L}, t)|^2}{|\vec{r}_2 - \vec{r}_2\!'|} \\
& = & - G m \int d \vec{r}_2\!' \ \frac{m|\psi_2(-\vec{r}_2\!'-\vec{L}, t)|^2}{|\vec{r}_2 - \vec{r}_2\!'|},
\end{eqnarray}
where in the first line we changed the integration variable to $\vec{r}_2\!'$, and in the second line we utilized the mirror symmetry and the independence of $\varphi$.
This way Eqs.~(\ref{EQ_COUPLED}) decouple and it is sufficient to solve only one single-particle equation:
\begin{equation}
i \hbar \pdv{t} \psi(\vec r,t) = 
\left( - \frac{\hbar^2}{2 m}\nabla^2 + V\qty[\abs{\psi}^2]
\right) \psi(\vec r,t)
\label{EQ_V_DECOUPLED}
\end{equation}
with the potential energy term at a given time given by
\begin{equation}
V = - G m^2
\int   d \vec{r} \,' \ \frac{1}{|\vec r - \vec r \,'|} \Big( |\psi(\vec r \, ',t)|^2 + |\psi(-\vec r \, '-\vec{L},t)|^2 \Big) .    
\end{equation}
An analogical $V$ is obtained for the first particle (replace $-\vec L \to \vec L$) and hence we dropped the index `2'.
This integral is further simplified in~\ref{appendix:SN-Decouple} to cater for future numerical implementations.

Interestingly, although the bipartite wave function retains its product form at all times, the system is evolving in such a way that the average dynamics follows the classical predictions, as expected from the Ehrenfest's theorem~\cite{Kalbermann_2004}. 
The derived single-particle equation provides a simple method to verify this.
Firstly, the mean momentum evolves as $\dv{t}\ev{\vec p} = - \frac{i}{\hbar} \langle [\vec p, H] \rangle$, because the only assumption to derive this relation is the hermiticity of the Hamiltonian. 
Secondly, one finds that the right-hand side is equal to the (negative) gradient of potential energy, which is evaluated as:
\begin{equation}
\dv{t}\ev{\vec p} = - G m^2
\int  d \vec r \  \abs{\psi(\vec r)}^2  \int   d \vec{r}\, ' \ \frac{\vec r - \vec r \,'}{\abs{\vec r - \vec r \,'}^3} \
\Big(  |\psi(\vec r \, ')|^2 +  |\psi(-\vec r \, '-\vec{L})|^2 \Big) .
\end{equation}
We recognize this as Newton's second law with gravitational force between two continuous mass distributions, one dependent on $\vec r$ and the other on $\vec r \, '$.
The first term in the parenthesis is the average force of self-interaction and it is equal to zero. 
This is because due to the symmetry in the double integral where any two points within the mass distribution are counted twice: when vector $\vec r$ aims at the first point and vector $\vec r \, '$ at the second and vice versa, so the two contributions cancel out. Thus:
\begin{equation}
\dv{t}\ev{\vec p} = - \ Gm^2 \int  d \vec r \ \abs{\psi(\vec r)}^2  
\int   d \vec{r}\, ' \ \frac{\vec r - \vec r \,'}{\abs{\vec r - \vec r \,'}^3} \  |\psi(-\vec r \, '-\vec{L})|^2 .
\end{equation}
If $L$ is larger than the size of the mass distribution, as typically expected in the laboratory, then the wave functions can be approximated by Dirac deltas and the right-hand side reduces to Newtonian gravity force between point particles a distance $L$ apart. 
Note that the completely local interaction terms $V_{jk}(\vec r_j)$, which depend on the position of one particle only, on average produce Newton's force that depends on the distance between two particles.

The mathematical expressions for SN potentials obtained with initial Gaussian states (see Appendices) are formally similar to expressions in semi-classical gravity sourced by continuous spontaneous localization~\cite{Tilloy_2016,Tilloy_2019,Khosla_2018}, albeit there the Gaussians appear as smearing functions for the measurements of mass density.
It was shown in Ref.~\cite{Khosla_2018} that Gaussian regularization removes divergence in time evolution of the kinetic energy and we, therefore, verify how the corresponding quantity behaves in the SN model.
In \ref{appendix:AverageKE} we calculate the rate of change of the kinetic energy and argue that it is indeed finite. The source of regularization in the SN model is the delocalization of the (square-integrable) wave functions.

\section{Testing the SN Model}

\begin{figure}
    \centering
    \includegraphics[width=\linewidth]{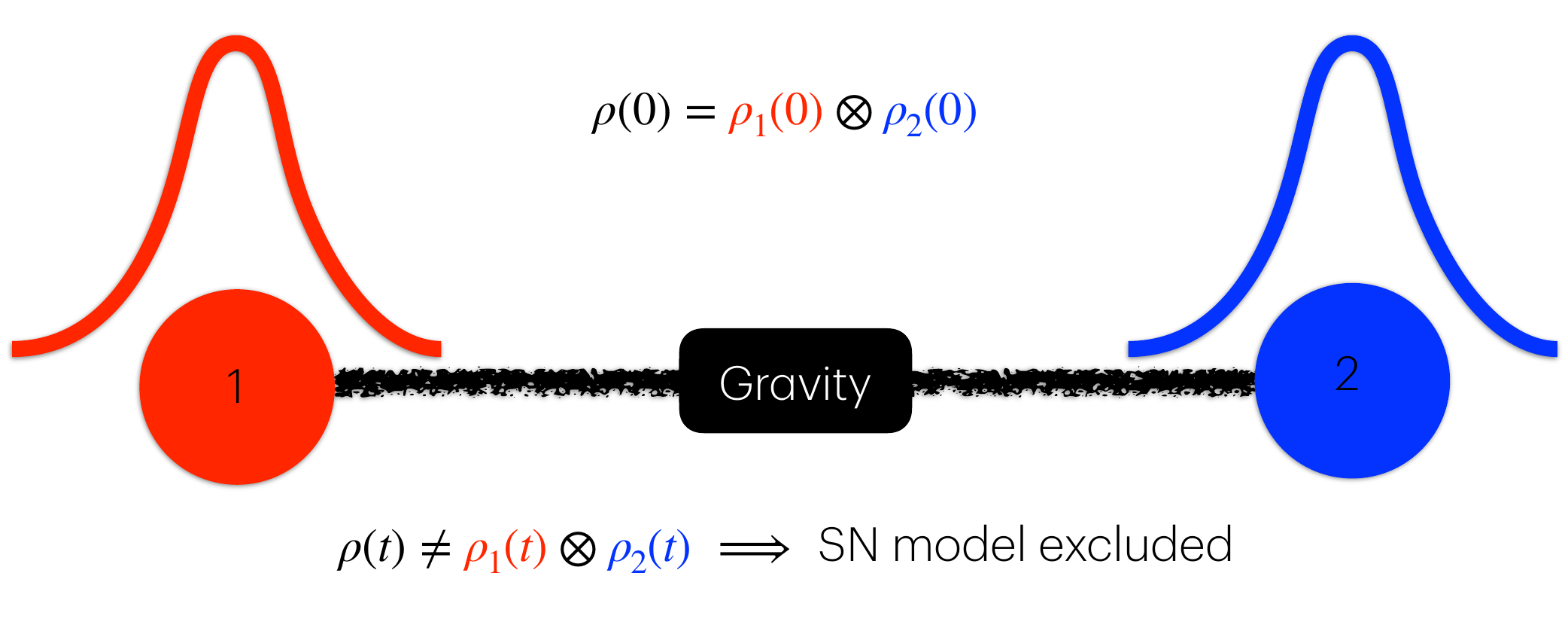}
    \caption{Schematic setup for testing the Schr\"odinger-Newton model. Two identical particles are initially independently prepared in Gaussian states by cooling in harmonic traps. The traps are thereafter opened, and the masses interact with each other during a free fall. In the main text we show that the SN model predicts that the initial product state remains product. Therefore, an observation of gravity-induced non-product states in a laboratory would disprove the model.
    Furthermore, in standard quantum mechanics the initial product state indeed becomes non-product and this occurs well before it becomes entangled.}
    \label{fig:Setup}
\end{figure}

\begin{figure}[!b]
\centering   
\includegraphics[width=\linewidth]{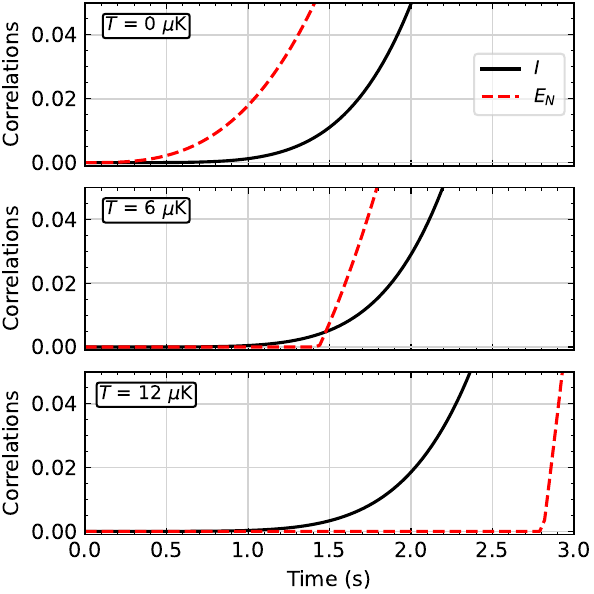}
\caption{Quantum entanglement (logarithmic negativity, $E_N$) and a measure of the non-product form of quantum states (quantum mutual information, $I$) for two Silica spheres coupled by quantum Newtonian potential. Each particle has a radius 100 nm  (mass $\approx 11.1$ fg). They are initially separated by a distance of 500 nm and individually cooled in harmonic traps of frequency $500$ kHz. The three plots show the time evolution of correlations for various temperatures. All single-particle parameters have been realized in Ref.~\cite{Aspelmeyer2020-Science} reaching temperature $T = 12 \ \mu$K (see the bottom panel). In realistic situations (of non-zero temperatures), the system evolves into non-product states well before it gets entangled. Therefore, testing the Schrödinger-Newton
model requires shorter coherence time than witnessing gravitational entanglement in the same setup.}
\label{fig:Correlations-QM}
\end{figure}

The results obtained allow a test of the SN model with the setups that have already been discussed for observation of gravitational entanglement~\cite{AlBalushi2018,Tanjung2020,Qvarfort2020,vandeKamp2020,Rijavec2021,Weiss2021,Carney2021,Kustura2022,Pedernales2022,Marshman2022,Christodoulou2023}, but with relaxed demands as we shall now explain.
First of all, according to the preceding section, the observation of quantum entanglement produced from uncorrelated states is incompatible with the SN model.
In the case of pure states, entanglement is identical to the non-product form of the quantum state, i.e., any state not in the form $\psi_1 \psi_2$ is entangled.
In practice, pure states are never produced due to unavoidable noise and finite temperature.
Instead, one deals with initial ensembles where the first mass occupies the state $\psi_1^{(j)}(0)$ with probability $p_j$, and the second mass independently occupies state $\psi_2^{(k)}(0)$ with probability $q_k$.
Due to this independence, the two systems are said to be uncorrelated with their density matrix in the form $\rho_1 \otimes \rho_2$.
According to the SN model, any product state evolves to a product state $\psi_1^{(j)}(0) \, \psi_2^{(k)}(0) \to \psi_1^{(j)}(t)  \,\psi_2^{(k)}(t)$ and the corresponding probabilities $p_j$ and $q_k$ do not change. Hence, initially uncorrelated ensembles evolve only into uncorrelated ensembles.
Therefore, the SN model would be excluded if one observes experimentally correlations established by gravity that are incompatible with any state of the form $\rho_1 \otimes \rho_2$.
See Fig.~\ref{fig:Setup} for a pictorial representation.
It follows that the class of states at variance with the SN model includes the majority of disentangled states, simplifying the experimental requirements.
In particular, such an experiment can be shorter in time.
In the next paragraph we present experimentally feasible parameters giving rise to this effect within the \emph{standard} quantum formalism.

In standard quantum theory, one of the ways to describe the gravitational coupling between non-relativistic point particles is the quantized Newton's potential, see e.g.~\cite{Bose2017,Vedral2017,Tanjung2020,Qvarfort2020,Ankit2023-QuGrav}.
In order to demonstrate the evolution towards non-product states, it is sufficient to consider one-dimensional motion.
In this case, the point masses attracting each other along the symmetry axis evolve under a one-dimensional equation:
\begin{equation}
i\hbar\pdv{t} \chi(z_1,z_2,t) 
 =  \left( -\frac{\hbar^2}{2m}  \pdv[2]{z_1} -\frac{\hbar^2}{2m} \pdv[2]{z_2} -  \frac{Gm^2}{ L+z_2-z_1 } \right)  \chi(z_1,z_2,t) .
\end{equation}
Note the key difference from the SN equation: in the Newtonian potential the two particles are directly coupled as the energy depends on $z_2 - z_1$.
Given that the displacements are much smaller than the separation, the potential can be expanded in a binomial series. It turns out that the cubic and higher-order terms are relevant only when there is significant relative motion between the two particles~\cite{Ankit2023-QuGrav}, and a truncation at the quadratic term is sufficient for practical purposes.
In this approximation, the initial product of Gaussian states,
\begin{equation}
\chi(z_1,z_2,t=0) 
=   \prod_{j=1,2} \frac{1}{(2 \pi \sigma^2)^{1/4}} \exp(- \frac{z_j^2}{4 \sigma^2}),
\end{equation}
evolves exclusively into Gaussian states.
This scenario, with two identical masses $m$ initially cooled to the ground states of harmonic traps with frequency $\omega_0$, was studied in Refs.~\cite{Tanjung2020,Qvarfort2020,Ankit2023-QuGrav}. 
The system is fully characterized by its covariance matrix, which was derived in an exact closed form in Ref.~\cite{Ankit2023-QuGrav} and for completeness is given in~\ref{appendix:CovarianceMatrix}.
We measure the entanglement by logarithmic negativity and the deviation from the product form is characterized by the quantum mutual information~\cite{Simon2000,Vidal2002,Adesso2004,Adesso2005}, see~\ref{appendix:QuantumCorrelations} for detailed expressions involving the covariance matrix.

In order to stay close to presently achievable parameters, we consider the experiment of Ref.~\cite{Aspelmeyer2020-Science}, which reported the cooling of a Silica nanoparticle down to a temperature of $12 \ \mu$K.
We assume that two identical Silica spheres of radius $R_0 = 100$ nm, separated by a distance $L=5 R_0$, are prepared by cooling them in harmonic traps of frequency 500 kHz.
For closed dynamics, the covariance matrix of the system, initialized in a product of the thermal states, equals $\bm{\sigma}_\text{th} (t) = (2\bar{n}+1)\bm{\sigma} (t)$, where $\bm{\sigma} (t)$ is the covariance matrix when starting from the product of the ground states and $\bar{n} = \qty[ \exp(\hbar \omega_0/k_B T) - 1 ]^{-1}$ represents the phonon number at temperature $T$.
Fig.~\ref{fig:Correlations-QM} shows logarithmic negativity and quantum mutual information dependence on time for various temperatures. 
While numerical values of logarithmic negativity and mutual information cannot be directly compared in general~\cite{Ganardi2022}, we can compare the regions where negativity is almost zero.
This is because we are considering two-mode Gaussian states, where logarithmic negativity detects any non-separability~\cite{Simon2000}.
The bottom panel is for the experimentally achieved temperature of $12 \ \mu$K and shows that considerable amounts of mutual information (which quantifies the distance to the set of all product states~\cite{Modi2010}) 
can be observed within one second in advance of entanglement.

\section{Signaling}
As a nonlinear extension of the Schr\"odinger equation, the SN dynamics can be exploited for superluminal signaling. We first review the argument demonstrating this, give a concrete example of signaling in the SN model, and finally propose a way out of this problem.

\begin{figure}
\centering
\includegraphics[width=0.7\linewidth]{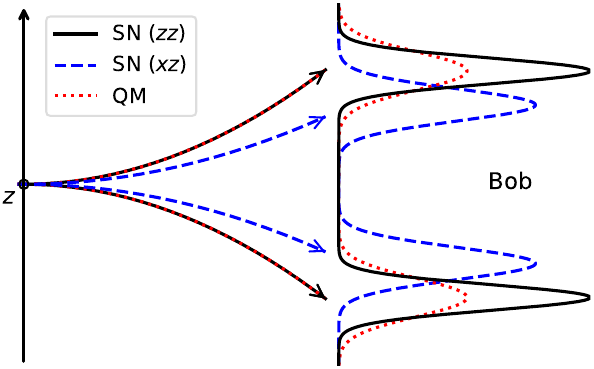}
\caption{Superluminal communication in the Schr\"odinger--Newton model, see also Ref.~\cite{Bahrami2014}.
Consider two massive spin-$\frac{1}{2}$ particles prepared in the singlet state $| \psi^- \rangle = (\ket{\uparrow \downarrow} - \ket{\downarrow \uparrow})/\sqrt{2}$.
Alice, who measures the first particle, encodes a binary value by choosing to perform a spin measurement (using Stern-Gerlach magnet) along the $z$ axis or the $x$ axis.
Bob, who measures the second particle, can read Alice's choice from the outcomes of his $z$ measurement.
The plot shows the distribution of Bob's particles for the two cases and compares them with the canonical quantum mechanical results.
If Alice measures along $z$, the spin state of the second particle is also along $z$.
While propagating towards the screen, self-gravity limits the spread of the wave function, producing a distribution (black) that is narrow in comparison with the quantum mechanical prediction (dotted red).
If Alice measures along $x$, the spin state of the second particle collapses along $x$, i.e., it is given by the quantum superposition of spin states along the $z$ axis.
In this case, self-gravity pulls the two components of the superposition towards the center between them, shifting the peaks of the dashed blue curves towards each other.
Finally, Bob reads Alice's bit from the position where his particles arrive the most often.}
\label{fig:signalling}
\end{figure}
Let us briefly review Gisin's argument~\cite{Gisin_1989}, see also Ref.~\cite{Bahrami2014} for a discussion on signaling specifically in the SN model.
Suppose a deterministic and nonlinear evolution of system $S$ is described by a map $g$, i.e., an initial state $|\psi \rangle \langle \psi |$ evolves to state $g(|\psi \rangle \langle \psi |)$.
If one starts with an ensemble $\Bqty{p_j, |\psi_j \rangle \langle \psi_j | }$, it evolves to $\Bqty{p_j, g(|\psi_j \rangle \langle \psi_j |) }$.
Consider now two ensembles, $\Bqty{p_j, |\psi_j \rangle \langle \psi_j |}$ and $\Bqty{q_k, |\phi_k \rangle \langle \phi_k |}$, which produce the same density matrix $\rho = \sum_j p_j |\psi_j \rangle \langle \psi_j | = \sum_k q_k |\phi_k \rangle \langle \phi_k |$.
Since $g$ is nonlinear, we can find two ensembles 
such that after the evolution $\sum_j p_j g(|\psi_j \rangle \langle \psi_j |) \neq \sum_k q_k g(|\phi_k \rangle \langle \phi_k |)$.
Hence, from differing final density matrices one can also distinguish the initial ensembles.
This leads to superluminal communication once we introduce another system, $R$, which purifies $\rho$.
It is known that any ensemble that forms a pure state decomposition of $\rho$ can be prepared via a measurement on $R$~\cite{Nielsen2012}.
In this way $R$ sends a message to $S$ by choosing the measurement to prepare $S$ either in ensemble $\Bqty{p_j, |\psi_j \rangle \langle \psi_j |}$ or ensemble $\Bqty{q_k, |\phi_k \rangle \langle \phi_k |}$.
$S$ then recovers the message by evolving the system and verifying whether it is described by $\sum_j p_j g(|\psi_j \rangle \langle \psi_j |)$ or $\sum_k q_k g(|\phi_k \rangle \langle \phi_k |)$.

The standard formulation of the SN model suffers from the problem just described. 
The origin is the nonlinear term representing self-gravity, for a single particle it is the term in Eq.~(\ref{EQ_1SN}).
Fig.~\ref{fig:signalling} shows explicitly how to exploit the above argument for superluminal communication.
One way to resolve this signaling issue is to extend the SN equation to mixed states.
Specifically, if a particle is described by a mixed state $\rho$, then any pure state within the ensemble evolves as:
\begin{equation}
i \hbar  \pdv{t} \psi_j(\vec{r}, t)
= \left(
- \frac{\hbar^2}{2m} \nabla^2
- G m \int d \vec{r'} \ \frac{m \rho(\vec r \, ', t)}{|\vec{r} - \vec r \,'|} \right) \psi_j(\vec{r}, t),
\label{EQ_MIXED_SN}
\end{equation}
i.e., the probability density $\rho(\vec r, t) = \langle \vec r | \rho_t | \vec r \rangle$ replaces $|\psi_j(\vec r, t)|^2$ in the usual formulation of the SN equation.
In this way, if two ensembles are initially compatible with the density, $\rho(\vec r, 0) = \sum_j p_j |\psi_j(\vec r, 0)|^2 = \sum_k q_k |\phi_k(\vec r, 0)|^2$, evolving each pure state according to Eq.~(\ref{EQ_MIXED_SN}) and taking the average reveals that they agree with the density at all times, i.e., $\rho(\vec r, t) = \sum_j p_j |\psi_j(\vec r, t)|^2 = \sum_k q_k |\phi_k(\vec r, t)|^2$.
In fact, the density matrix $\rho_t$
satisfies the von Neumann equation:
\begin{equation}
i \hbar \pdv{t}\rho_t = [H(\rho_t),\rho_t],
\label{EQ_MIXED_VN}
\end{equation}
where the new Hamiltonian depends only on the ensemble average $\rho_t$:
\begin{equation}
    H(\rho_t) = - \frac{\hbar^2}{2m} \nabla^2
- G m \int d \vec{r}\,' \ \frac{m \rho(\vec r \, ', t)}{|\vec{r} - \vec r \,'|} .
\end{equation}
Since the commutator is independent of the wave function $\psi(\vec r,t)$ and solely dependent on the density $\rho(\vec r, t)$, all ensembles that conform to $\rho(\vec r, t)$ are indistinguishable at any time and therefore cannot be exploited for signaling.

In order to compute time evolution of the system according to the modified SN equation, one can proceed in two ways. The first one is to solve the nonlinear von Neumann equation~(\ref{EQ_MIXED_VN}) to obtain $\rho_t$, which is then used in the SN equation~(\ref{EQ_MIXED_SN}) to determine the evolution of any given state $\psi_j(\vec r, t)$.
Alternatively, one can use only the modified SN equation.
Let us demonstrate the computer code based on this approach.
It begins with the initial condition given by the whole ensemble $\qty{ p_j, \psi_j(\vec r, 0) }$, from which one computes $\rho(\vec r, 0)$. Then equation~(\ref{EQ_MIXED_SN}) is integrated for an infinitesimal step $\Delta t$ for every element of the initial ensemble, leading to $\qty{ p_j, \psi_j(\vec r, \Delta t) }$, from which one computes $\rho(\vec r, \Delta t)$, and so on.

We qualitatively illustrate this approach by solving the issue shown in Fig.~\ref{fig:signalling}. 
The state of the second particle, sent to Bob, is completely mixed with probability density $\rho(\vec{r}, t) = \frac{1}{2} |\psi_\uparrow(\vec{r}, t)|^2 + \frac{1}{2} |\psi_\downarrow(\vec{r}, t)|^2$, written using Bob's measurement basis. 
If a spin up (spin down) particle propagates according to Eq.~(\ref{EQ_MIXED_SN}), with $\rho(\vec{r}, t)$ embodied in the potential energy, its wave function is narrowed (shifted upwards) due to the interaction described by $|\psi_\uparrow(\vec{r}, t)|^2$ and it is being shifted downwards (narrowed) due to the interaction described by $|\psi_\downarrow(\vec{r}, t)|^2$. The net result is that both peaks are narrowed and pulled towards each other as compared to the canonical quantum predictions, i.e. they now overlap with the dashed blue curves in Fig.~\ref{fig:signalling}.
Note that these shifts and widths are, in principle, subject to experimental tests. However, the expected discrepancies are of the same order as those in the single-particle SN dynamics, and hence, are hard to observe.

The modification of Eq.~(\ref{EQ_MIXED_SN}) provides a straightforward solution to the signaling problem of the SN model.
Note that this modification is still deterministic and nonlinear.
It is therefore interesting to ask how exactly it escapes the argument by Gisin.
The answer is that the dynamical map corresponding to Eq.~(\ref{EQ_MIXED_SN}) is a function of \emph{both} the pure state and the density matrix whereas the original argument additionally assumed that the map $g$ is a function of the pure state only.

Yet, one should verify if the modification is consistent in all physical situations.
The signaling problem forces the use of Eq.~(\ref{EQ_MIXED_SN}) for subsystems of entangled systems (improper mixtures).
Additionally, before any measurements are conducted, there is no way of favoring any particular ensemble compatible with the reduced density matrix, implying that the modification is a rather natural approach. 
One way to retain the no-signaling is to evolve the local subsystem according to Eq.~(\ref{EQ_MIXED_SN}), even when the distant particle is measured, and to modify the local density matrix only upon acting on the local particle.
Consequently, proper mixtures of individual particles (coming from sources that emit single quantum systems in a particular pure state with a particular probability) would evolve under the original pure-state SN equation.
As an additional supporting argument, note that the particle emitted in a particular experimental run has no information about the wave functions of particles that will be emitted in the future and, hence, about the statistical ensemble, thereby forbidding calculations with Eq.~(\ref{EQ_MIXED_SN}).
On the other hand, one may argue that in practice any source of single particles admits a density matrix, possibly changing over time, allowing for the use of Eq.~(\ref{EQ_MIXED_SN}).
Accordingly, there exist apparently consistent options for the dynamical maps which depend on both the pure state and the density matrix, and it is a matter of experiment to disqualify or distinguish between them.
For example, to decide between the pure-state SN equation and the mixed-state modification (\ref{EQ_MIXED_SN}) a source could send an even mixture of particles with spin-up and spin-down along the $z$ direction, followed by the $z$-axis Stern-Gerlach measurement. The pure-state SN equation predicts the pattern shown by the black lines in Fig.~\ref{fig:signalling}, whereas the mixed-state SN equation predicts the dashed line pattern.

\section{Conclusions}
We argued that the Schr\"odinger--Newton model restricts the initial product states to evolve only into product states.
Yet, on average one recovers classical mechanics of continuous mass distributions.
This is very different from the predictions of canonical quantum mechanics, where the same initial conditions lead to various types of correlations, culminating in quantum entanglement.
Assuming currently accessible parameters, we computed the standard evolution into non-product states due to the quantum Newtonian potential, and demonstrated that testing the Schrödinger-Newton model requires shorter coherence time than witnessing gravitational entanglement in the same setup.
We also showed how to extend the SN model to mixed states in a way that avoids problems with superluminal communication.
If an experiment confirms the existence of a gravity-induced non-product (mixed) quantum state, the SN equation will have to be rejected.

\ack
This work is jointly supported by the 
(i) Polish National Agency for Academic Exchange NAWA, project no. PPN/PPO/2018/1/00007/U/00001, 
(ii) Xiamen University Malaysia, project no. XMUMRF/2022-C10/IPHY/0002,
and (iii) Fulbright Senior Award granted by the Polish-U.S. Fulbright Commission.
K.K. was supported by NCN project SONATA-BIS no. 2019/34/E/ST1/00390.
A.K. is supported by the QuantERA II project ``ExTRaQT'' contract no. 2021/03/Y/ST2/00178, that has received funding from EU’s Horizon 2020, and NCN SONATA-BIS grant No. 2017/26/E/ST2/01008.
P.A. acknowledges the support from SERB-DST, Govt.~of India, via project no. CRG/2022/009359.
We thank Andr\'{e} H.A. Malavazi (ICTQT, Poland) and Gagandeep Singh for their comments on the initial draft of this manuscript.
We acknowledge National Supercomputing Mission (NSM) for providing computing resources of `PARAM Ganga' at IIT Roorkee, which is implemented by C-DAC and supported by the MeitY and DST, Govt.~of India.
The authors gratefully thank the anonymous reviewers for their comments and suggestions which have significantly improved the presentation of this article.

\appendix

\section{SN Potential at $t=0$}
\label{appendix:SN-Pot-InitialTime}

Here we provide closed-form expressions for the potential energies $V_{jk}$ defined in Eqs.~\eqref{pot-self} and~\eqref{pot-mutual}, assuming initial Gaussian states of the masses, and emphasize that they are functions of either $\vec r_1$ or $\vec r_2$. This allows us to write the total Hamiltonian as a sum of commuting (local) terms.
At time $t=0$ the two masses are independently prepared in symmetric 3D Gaussian wave packets and the state describing them is $\Psi(\vec{r}_1,\vec{r}_2) = \psi_1(\vec{r}_1)\psi_2(\vec{r}_2)$, where
\begin{equation}
    \psi_j(\vec{r}_j) = \frac{1}{(2\pi\sigma^{2})^{3/4}}\exp\bigg(-\frac{r_{j}^{2}}{4\sigma^{2}}\bigg) .
  \label{eq_InitialGaussian}
\end{equation}
The self-interaction term $V_{11}$ is given as
\begin{equation}
V_{11}  =  -G m \int d\vec{r}_1\!'d\vec{r}_2\!'\frac{m|\Psi(\vec{r}_1\!',\vec{r}_2\!')|^{2}}{|\vec{r}_{1}-\vec{r}_1\!'|}
= -Gm^{2}\int d\vec{r}_1\!'\frac{|\psi_{1}(\vec{r}_1\!')|^{2}}{|\vec{r}_{1}-\vec{r}_1\!'|},
\end{equation}
where we utilized the normalization $\int d\vec{r}_2\!'|\psi_{2}(\vec{r}_2\!')|^{2} = 1$.
Assuming the $z$-axis is oriented along $\vec r_1$, one can perform the angular integration by standard means:
\begin{equation}
\int \frac{d\Omega}{|\vec{r}-\vec{r}\,'|}=\frac{4\pi}{r_>} ,
\end{equation}
where $r_> = \max(r,r')$, which implies
\begin{align}
V_{11} =& -Gm^{2}\bigg(\int_{0}^{r_{1}}dr_{1}'\ r_{1}'^{2}|\psi_{1}(r_{1}')|^{2}\frac{4\pi}{r_{1}}+\int_{r_{1}}^{\infty}dr_{1}'\ r_{1}'^{2}|\psi_{1}(r_{1}')|^{2}\frac{4\pi}{r_{1}'}\bigg)
\\
=& -4\pi Gm^{2}\bigg(\frac{1}{r_{1}}\int_{0}^{r_{1}}dr_{1}'\ r_{1}'^{2}|\psi_{1}(r_{1}')|^{2}+\int_{r_{1}}^{\infty}dr_{1}'\ r_{1}'|\psi_{1}(r_{1}')|^{2}\bigg)
 \nonumber\\
=& -\frac{4\pi Gm^{2}}{(2\pi\sigma^{2})^{3/2}}\Bigg[\frac{1}{r_{1}}\int_{0}^{r_{1}}dr_{1}'\ r_{1}'^{2}\exp\bigg(-\frac{r_{1}'^{2}}{2\sigma^{2}}\bigg)+\int_{r_{1}}^{\infty}dr_{1}'\ r_{1}'\exp\bigg(-\frac{r_{1}'^{2}}{2\sigma^{2}}\bigg)\Bigg]
 \nonumber\\
=& -\frac{4\pi Gm^{2}}{(2\pi\sigma^{2})^{3/2}}\Bigg[\frac{1}{r_{1}}(\sigma\sqrt{2})^{3}\int_{0}^{r_{1}/\sigma\sqrt{2}}d\zeta\ \zeta^{2}e^{-\zeta^{2}}+(\sigma\sqrt{2})^{2}\int_{r_{1}/\sigma\sqrt{2}}^{\infty}d\zeta\ \zeta
 e^{-\zeta^{2}}\Bigg] .
 \nonumber
\end{align}
The two integrals can be evaluated as:
\begin{align}
\int_{0}^{r_{1}/\sigma\sqrt{2}}d\zeta\ \zeta^{2}e^{-\zeta^{2}} =& \qty[ \dfrac{\sqrt{\pi} }{4} \ \text{erf}\left(\zeta\right)-\dfrac{\zeta\mathrm{e}^{-\zeta^{2}}}{2} ]_{0}^{r_{1}/\sigma\sqrt{2}}
\nonumber\\
=&\frac{\sqrt{\pi}}{4}\mathrm{\ \text{erf}}\left(\frac{r_{1}}{\sigma\sqrt{2}}\right)-\frac{r_{1}}{2\sigma\sqrt{2}}\exp\bigg(-\frac{r_{1}^{2}}{2\sigma^{2}}\bigg) ,
   \\[0.5em]
\int_{r_{1}/\sigma\sqrt{2}}^{\infty}d\zeta\ \zeta e^{-\zeta^{2}} =& -\frac{1}{2} \qty[e^{-\zeta^{2}} ]_{r_{1}/\sigma\sqrt{2}}^{\infty}=\frac{1}{2}\exp\bigg(-\frac{r_{1}^{2}}{2\sigma^{2}}\bigg) ,
\end{align}
where the error function is defined as
\begin{equation}
\text{erf}(z) = \frac{2}{\sqrt{\pi}} \int_{0}^{z} dt \ e^{-t^2} .
\end{equation}
This leads to the following expression for the potential:
\begin{equation}
V_{11}(\vec{r}_{1})  = 
-\frac{Gm^{2}}{r_{1}}\mathrm{\ erf}\left(\frac{r_{1}}{\sigma\sqrt{2}}\right) ,
\end{equation}
and by extension,
\begin{equation}
 V_{22}(\vec{r}_{2})  = 
-\frac{Gm^{2}}{r_{2}}\mathrm{\ erf}\left(\frac{r_{2}}{\sigma\sqrt{2}}\right) .
\end{equation}
Similarly, the mutual-interaction term $V_{21}$ can be written as
\begin{align}
V_{21}  =& -Gm \int d\vec{r}_1\!'d\vec{r}_2\!'\frac{m|\Psi(\vec{r}_1\!',\vec{r}_2\!')|^{2}}{|\vec{r}_{2}-\vec{r}_1\!'+\vec{L}|} 
\\
=& - Gm^{2} \int d\vec{r}_1\!'\frac{|\psi_{1}(\vec{r}_1\!')|^{2}}{|\vec{r}_{2}-\vec{r}_1\!'+\vec{L}|} \int d\vec{r}_2\!' |\psi_{2}(\vec{r}_2\!')|^{2}
\nonumber  \\
=& -Gm^{2}\bigg(\int_{0}^{|\vec{r}_{2}+\vec{L}|}dr_{1}'\ r_{1}'^{2}|\psi_{1}(r_{1}')|^{2}\frac{4\pi}{|\vec{r}_{2}+\vec{L}|}+\int_{|\vec{r}_{2}+\vec{L}|}^{\infty}dr_{1}'\ r_{1}'^{2}|\psi_{1}(r_{1}')|^{2}\frac{4\pi}{r_{1}'}\bigg),
\nonumber
\end{align}
which has the same mathematical form as $V_{11}$, except that $r_{1}$ is replaced by $|\vec{r}_{2}+\vec{L}|$. Hence,
\begin{equation}
V_{21}(\vec{r}_{2}) =  -\frac{Gm^{2}}{|\vec{r}_{2}+\vec{L}|}\mathrm{\ erf}\left(\frac{|\vec{r}_{2}+\vec{L}|}{\sigma\sqrt{2}}\right) ,
\end{equation}
and by extension
\begin{equation}
V_{12}(\vec{r}_{1})  =  -\frac{Gm^{2}}{|\vec{r}_{1}-\vec{L}|}\mathrm{\ erf}\left(\frac{|\vec{r}_{1}-\vec{L}|}{\sigma\sqrt{2}}\right)  .
\end{equation}
Finally, at $t=0$ we have explicit separation into terms dependent only on $\vec r_1$ and only on $\vec r_2$: $H(\vec{r}_{1},\vec{r}_{2},0) = H_1(\vec{r}_{1},0) + H_2(\vec{r}_{2},0)$, where
\begin{align}
H_1(\vec{r}_{1},0) =& 
- \frac{\hbar^2}{2m} \nabla_1^2 -\frac{Gm^{2}}{r_{1}}\mathrm{\ erf}\left(\frac{r_{1}}{\sigma\sqrt{2}}\right) -\frac{Gm^{2}}{|\vec{r}_{1}-\vec{L}|}\mathrm{\ erf}\left(\frac{|\vec{r}_{1}-\vec{L}|}{\sigma\sqrt{2}}\right) ,
\nonumber   \\[0.5em]
H_2(\vec{r}_{2},0) =&
- \frac{\hbar^2}{2m} \nabla_2^2 -\frac{Gm^{2}}{r_{2}}\mathrm{\ erf}\left(\frac{r_{2}}{\sigma\sqrt{2}}\right) -\frac{Gm^{2}}{|\vec{r}_{2}+\vec{L}|}\mathrm{\ erf}\left(\frac{|\vec{r}_{2}+\vec{L}|}{\sigma\sqrt{2}}\right) .
\label{EQ_H1+H2}
\end{align}

\section{Preservation of the product form}
\label{appendix:SN-ProductForm-Proof}

Assume that initially the two masses are in a product state 
$\Psi(\vec r_1, \vec r_2, 0)=\psi_{1}(\vec r_1, 0)\psi_{2}(\vec r_2, 0)$, and that the total Hamiltonian is a sum of commuting parts $H(\vec{r}_1,\vec{r}_2,0)=H_{1}(\vec{r}_1,0) + H_{2}(\vec{r}_2,0)$, see Eq.~\eqref{EQ_H1+H2} for a concrete example, where $[H_{1}(\vec{r}_1,0), H_{2}(\vec{r}_2,0)] = 0$.
We will use the abbreviation $ f_j(\vec{r}_j) = f_j$.
Given that the time-evolution is smooth and continuous, the bipartite wave function after an infinitesimal time step $\Delta t$ can be expanded in a Taylor's series in $\Delta t$:
\begin{equation}
\Psi(\Delta t)=\sum_{n=0}^{\infty}\frac{(\Delta t)^{n}}{n!} \partial_t^n \Psi,
\end{equation}
where the derivatives on the right-hand side are evaluated at $t=0$. 
The bipartite wave function stays in a product form if and only if
\begin{align}
\sum_{n=0}^{\infty} \frac{(\Delta t)^{n}}{n!}\partial_{t}^{n}\Psi
=& \left(\sum_{k=0}^{\infty}\frac{(\Delta t)^{k}}{k!}
\partial_{t}^{k}\psi_{1}\right)\left(\sum_{\ell=0}^{\infty}\frac{(\Delta t)^{\ell}}{\ell!}\partial_{t}^{\ell}\psi_{2}\right)
\nonumber\\
 =&  \sum_{n=0}^\infty (\Delta t)^{n}\sum_{k=0}^{n}\frac{1}{k!(n-k)! } \ \partial_{t}^{k}\psi_{1} \ \partial_{t}^{n-k}\psi_{2} ,
 \hspace{.5cm}
 : \qty{n=k+\ell},
\nonumber\\
 =&  \sum_{n=0}^\infty \frac{(\Delta t)^{n}}{n!}\sum_{k=0}^{n}{n \choose k} \ \partial_{t}^{k}\psi_{1} \ \partial_{t}^{n-k}\psi_{2},
 \label{eq:Hypothesis}
\end{align}
i.e., when the first mass evolves solely under Hamiltonian $H_1$, and the second mass evolves solely under $H_2$.
We prove this statement by induction in the powers of $\Delta t$.
Assume that the hypothesis holds for the expression multiplying $(\Delta t)^n/n!$ on both sides:
\begin{equation}
\partial_{t}^{n}\Psi
= \sum_{k=0}^{n}{n \choose k} \ \partial_{t}^{k}\psi_{1} \ \partial_{t}^{n-k}\psi_{2} .
\end{equation}
The next time derivative is $\partial_{t}^{n+1}\Psi = \partial_t ( \partial_{t}^{n}\Psi )$ and by utilizing the inductive assumption we write:
\begin{align}
\partial_{t}^{n+1}\Psi
=& \partial_t \qty( \sum_{k=0}^{n}{n \choose k} \ \partial_{t}^{k}\psi_{1} \ \partial_{t}^{n-k}\psi_{2} )    
= \sum_{k=0}^{n}{n \choose k}\left(\partial_{t}^{k}\psi_{1} \ \partial_{t}^{n-k+1}\psi_{2}+\partial_{t}^{k+1}\psi_{1} \ \partial_{t}^{n-k}\psi_{2}\right)
\nonumber\\[0.5em]
=& {n \choose 0}\left(\psi_{1} \ \partial_{t}^{n+1}\psi_{2}+\partial_{t}\psi_{1} \ \partial_{t}^{n}\psi_{2}\right)+{n \choose 1}\left(\partial_{t}\psi_{1} \ \partial_{t}^{n}\psi_{2}+\partial_{t}^{2}\psi_{1} \ \partial_{t}^{n-1}\psi_{2}\right)
\nonumber\\
& +{n \choose 2}\left(\partial_{t}^{2}\psi_{1} \ \partial_{t}^{n-1}\psi_{2} + \partial_{t}^{3}\psi_{1} \ \partial_{t}^{n-2}\psi_{2}\right)+\dots+{n \choose n}\left(\partial_{t}^{n}\psi_{1} \ \partial_{t}\psi_{2} + \partial_{t}^{n+1}\psi_{1} \ \psi_{2}\right)
\nonumber\\[0.5em]
=& {n \choose 0} \ \psi_{1} \ \partial_{t}^{n+1}\psi_{2}+\left({n \choose 0} + {n \choose 1}\right)\partial_{t}\psi_{1} \ \partial_{t}^{n}\psi_{2} 
\nonumber\\
& + \left({n \choose 1}+{n \choose 2}\right) \ \partial_{t}^{2} \psi_{1} \ \partial_{t}^{n-1}\psi_{2}+\dots+{n \choose n} \ \partial_{t}^{n+1}\psi_{1} \ \psi_{2}
\nonumber\\[0.5em]
=& {n+1 \choose 0}  \ \psi_{1} \ \partial_{t}^{n+1}\psi_{2} + {n+1 \choose 1} \ \partial_{t}\psi_{1} \ \partial_{t}^{n}\psi_{2} + {n+1 \choose 2} \ \partial_{t}^{2}\psi_{1} \ \partial_{t}^{n-1}\psi_{2} 
\nonumber\\
&+ \dots +{n+1 \choose n+1} \ \partial_{t}^{n+1}\psi_{1} \ \psi_{2},
\hspace{5mm} :\qty{ {n \choose k}+{n \choose k+1}={n+1 \choose k+1} },
\nonumber\\[0.5em]
=& \sum_{k=0}^{n+1}{n+1 \choose k} \ \partial_{t}^{k}\psi_{1} \ \partial_{t}^{n-k+1}\psi_{2} .
\end{align}
Accordingly, if the expansion holds for the $n^\text{th}$ term, it also holds for the $(n+1)^\text{th}$ term, and, by extension, for all terms that follow thereafter. Note that the proof utilized only the properties of time derivative and the fact that suitable derivatives are multiplied by binomial coefficients.

To complete the proof, we now show that the inductive hypothesis holds for small $n$, where it becomes clear that the binomial coefficients follow from the commutativity of $H_1$ and $H_2$.
For $n=0$ the hypothesis is just our assumption about the initial state $\Psi(0) = \psi_1(0) \psi_2(0)$.
The case of $n=1$ follows from the preceding section, $H(0)=H_{1}(0)+H_{2}(0)$:
\begin{align}
\partial_{t}\Psi &= -\frac{i}{\hbar} H\Psi
\nonumber\\
&= -\frac{i}{\hbar} (H_{1}+H_{2})\psi_{1}\psi_{2}
\nonumber\\
&= -\frac{i}{\hbar}
\big( H_{1}\psi_{1} \ \psi_{2} + \psi_{1} \ H_{2}\psi_{2} \big)
\nonumber\\
&= \partial_{t} \psi_{1} \ \psi_{2} + \psi_{1} \ \partial_{t}\psi_{2} ,
\end{align}
and by the inductive hypothesis it holds for all the terms that follow.
Nevertheless, let us also explicitly compute the $n=2$ case, where the commutativity of $H_1$ and $H_2$ is readily apparent. On the one hand:
\begin{align}
\partial_{t}^2 \Psi
&=  \partial_{t} (\partial_{t} \Psi) 
\\
&= -\frac{i}{\hbar} \partial_t( H\Psi) 
\nonumber\\
&= -\frac{i}{\hbar} \Big( \qty( \partial_t H ) \Psi +  H  ( \partial_t \Psi)  \Big)
\nonumber\\
&= \left(-\frac{i}{\hbar}\right)( \partial_t H_{1}) \psi_1 \psi_2
+ \left(-\frac{i}{\hbar}\right) \psi_1 ( \partial_t H_{2}) \psi_2
+ \left(-\frac{i}{\hbar}\right)^2 (H_1 + H_2) (H_1 + H_2) \psi_1 \psi_2 ,
\nonumber
\end{align}
and on the other hand:
\begin{align}
(\partial_{t}^2 \psi_1) \psi_2 & +  2 (\partial_{t} \psi_1) (\partial_{t} \psi_2) + \psi_1 (\partial_{t}^2 \psi_2) 
\nonumber \\
=& 
\left(-\frac{i}{\hbar}\right) \partial_t (H_1 \psi_1) \psi_2
+ 2 \left(-\frac{i}{\hbar}\right)^2 (H_1 \psi_1) (H_2 \psi_2)
+ \left(-\frac{i}{\hbar}\right) \psi_1 \partial_t (H_2 \psi_2)
\nonumber\\
=&  \left(-\frac{i}{\hbar}\right)( \partial_t H_{1}) \psi_1 \, \psi_2
+ \left(-\frac{i}{\hbar}\right)^2 H_{1}^2 \psi_1 \, \psi_2
+ 2 \left(-\frac{i}{\hbar}\right)^2 H_1 H_2 \psi_1 \psi_2
\nonumber\\
& + \left(-\frac{i}{\hbar}\right) \psi_1 \, ( \partial_t H_{2}) \psi_2
+ \left(-\frac{i}{\hbar}\right)^2 \psi_1 H_2^2 \psi_2 . 
\end{align}
Both sides are equal if $H_1 H_2 + H_2 H_1$ (present on the left-hand side) is given by $2 H_1 H_2$ (present on the right-hand side), which is indeed satisfied for commuting $H_1$ and $H_2$.
All other expansion terms are then obtained by induction.
Accordingly, for an infinitesimal time step $\Delta t$ we have $\Psi(\Delta t) = \psi_1(\Delta t) \psi_2(\Delta t)$, and in continuous time evolution one concludes
$\psi_1(0) \psi_2(0) \to \psi_1(t) \psi_2(t) .$

The result above applies to arbitrary many-body systems. To see this, consider the SN Hamiltonian of $N$ particles, each of mass $m$, with potential $V = \sum_{j=1}^{N} \sum_{k=1}^{N} V_{jk}$, where:
\begin{equation}
V_{jk}(t)
 = - G m^2  \sum_{j=1}^{N} \sum_{k=1}^{N} 
  \int d\vec{r}_1\!' d\vec{r}_2\!' \dots d\vec{r}_N\!\!\!' \  \frac{|\Psi(\vec{r}_1\!',\vec{r}_2\!',\dots,\vec{r}_N\!\!\!',t)|^2}{|\vec{r}_j-\vec{r}_k\!'|}.   
\end{equation}
The total potential experienced by the $j^\text{th}$ particle is $\sum_{k=1}^{N} V_{jk}$, where $V_{jj}$ is the self-gravitation, and $V_{jk}$, for $j \neq k$, is its coupling with other masses. For an initial product state $\Psi(\vec{r}_1,\vec{r}_2,\dots,\vec{r}_N,0) = \psi_1(\vec{r}_1,0) \psi_2(\vec{r}_2,0) \dots \psi_N(\vec{r}_N,0)$ the potential at $t=0$ reads:
\begin{align}
V_{jk}
=& - G m^2  
\int d\vec{r}_1\!' d\vec{r}_2\!' \dots d\vec{r}_N\!\!\!' \ \frac{|\psi_1(\vec{r}_1\!',0)|^2 \ |\psi_2(\vec{r}_2\!',0)|^2 \ \dots \ |\psi_N(\vec{r}_N\!\!\!',0)|^2}{|\vec{r}_j-\vec{r}_k\!'|}
\nonumber\\
 =& - G m^2  
    \int d\vec{r}_1\!' |\psi_1(\vec{r}_1\!',0)|^2 
\    \dots  \
\int d\vec{r}_k\!' \frac{|\psi_k(\vec{r}_k\!',0)|^2}{|\vec{r}_j-\vec{r}_k\!'|}
\    \dots \
\int d\vec{r}_N\!\!' |\psi_N(\vec{r}_N\!\!\!',0)|^2
\nonumber\\
 =& - G m^2  
    \int d\vec{r}_k\!' \frac{|\psi_k(\vec{r}_k\!',0)|^2}{|\vec{r}_j-\vec{r}_k\!'|} ,
\hspace{1cm} : \qty{ \int d\vec{r}_m\!\!' |\psi_m(\vec{r}_m\!\!',0)|^2 = 1 } 
.
\end{align}
It is local as it depends on one coordinate only: $V_{jk} = V_{jk}(\vec{r}_j)$. For the case of identical initial Gaussian states of width $\sigma$,
\begin{equation}
	\psi(\vec r_j,t=0) = \frac{1}{(2\pi\sigma^2)^{3/4}} \exp\qty( - \frac{(\vec r_j-\ev{\vec r_j})^2}{4\sigma^2} ) ,
\end{equation}
the exact mathematical form is given by
\begin{equation}
V_{jk} = 
 -Gm^2
 \sum_{k=1}^{N}
  \frac{1}{|\vec{r}_{j}-\ev{\vec{r}_k}|}\mathrm{\ erf}\left(\frac{|\vec{r}_{j}-\ev{\vec{r}_k}|}{\sigma\sqrt{2}}\right) .
\end{equation}
Therefore, the full SN Hamiltonian at initial time is $H(t=0) = \sum_{j=1}^N  H_j(\vec{r}_j,t=0)$,
where the local terms read:
\begin{equation}
 H_j(\vec{r}_j,t=0)   =   -\frac{\hbar^2}{2m} \nabla^2_j -Gm^2
 \sum_{k=1}^{N}
  \frac{1}{|\vec{r}_{j}-\ev{\vec{r}_k}|}\mathrm{\ erf}\left(\frac{|\vec{r}_{j}-\ev{\vec{r}_k}|}{\sigma\sqrt{2}}\right).
\end{equation}
Note that all of the local Hamiltonians pairwise commute: $[H_j,H_k] = 0\ \forall (j,k)$.
We now use the argument above for different bipartitions to show that an initial product state evolves solely into a product state also in the multipartite case.

For simplicity we first consider a tripartite system.
The commutativity of local Hamiltonians implies that 
$[H_1,H_2 + H_3] = 0$, $[H_2,H_3 + H_1] = 0$, and $[H_3,H_2 + H_1] = 0$. As shown above, an initial product state $\Psi(0) = \psi_1(\vec{r}_1,0) \psi_2(\vec{r}_2,0) \psi_3(\vec{r}_3,0)$ evolves in three different partitions as 
\begin{equation}
\Psi(t) = \psi_1(\vec{r}_1,t)  \psi_{23}(\vec{r}_2,\vec{r}_3,t) = \psi_2(\vec{r}_2,t) \psi_{31}(\vec{r}_3,\vec{r}_1,t) = \psi_3(\vec{r}_3,t) \psi_{12}(\vec{r}_1,\vec{r}_2,t) ,    
\end{equation}
which is possible only when each and every particle evolves into a pure state: $\Psi(t) = \psi_1(\vec{r}_1,t) \psi_2(\vec{r}_2,t) \psi_3(\vec{r}_3,t)$. 
By extension, for a system of $N$ particles
\begin{equation}
\prod_{j=1}^{N} \psi_j(\vec{r}_j,0) \to \prod_{j=1}^{N} \psi_j(\vec{r}_j,t)
.
\end{equation}

\section{Simplifying the decoupled potential energy}
\label{appendix:SN-Decouple}

Here we utilize the elliptic integrals to simplify Eq.~(\ref{EQ_V_DECOUPLED}) in the main text. For completeness it is repeated here:
\begin{equation}
i \hbar \pdv{t} \psi(\vec r,t) = 
\left( - \frac{\hbar^2}{2 m}\nabla^2 + V\qty[\abs{\psi}^2]
\right) \psi(\vec r,t),
\end{equation}
where
\begin{equation}
V = - G m^2 
\int   d \vec{r} \,' \ \frac{1}{|\vec r - \vec r \, '|} \Big(  |\psi(\vec r \, ',t)|^2 +  |\psi(- \vec r \, '-\vec{L},t)|^2 \Big) .
\end{equation}
Consider the integral $I = \int d \vec{r} \, ' |\psi(\vec r \, ', t)|^2 / |\vec r - \vec r \, '|$ written in cylindrical coordinates:
\begin{equation}
I = 	\int s' ds' dz' d \phi' \frac{|\psi(s',z', t)|^2 }{\sqrt{s^2 + {s'}^2 - 2 s s' \cos(\phi - \phi') + (z-z')^2}}.
\label{eq:PHI_INT_Self}
\end{equation}
To simplify the notation we write the angular integral as 
\begin{equation}
\mathcal{K}  = \int_0^{2 \pi} \frac{d \phi'}{ \sqrt{a - b \cos(\phi - \phi')} } ,    
\end{equation}
where $	a  =  s^2 + {s'}^2 + (z - z')^2$ and $b  =  2 s s'$.
Given that the integration is over the full period of the cosine function, it is independent of the initial offset $\phi$.
Therefore,
\begin{align}
\mathcal{K}  &=  \int_0^{2 \pi} \frac{d \phi'}{\sqrt{a - b \cos \phi'}} 
\nonumber\\
&=  \int_0^{2 \pi} \frac{d \phi'}{ \sqrt{ a - b \qty(1 - 2\sin^2(\phi'/2) ) } } 
\nonumber\\
&= \frac{2}{\sqrt{a-b}} \int_0^{\pi} \frac{d \theta}{\sqrt{1 + \frac{2b}{a - b} \sin^2 \theta}}. 
\end{align}
One now recognizes that $\mathcal{K}$ is given by Eq.~(282.00) of Ref.~\cite{book-EllipInt-Byrd} with the following substitutions: 
$n^2 = 2b/(a-b)$, $k^2 = n^2/(1 + n^2) = 2b / (a+b)$, and $k' = \sqrt{1 - k^2} = \sqrt{(a-b)/(a+b)}$, where $n^2$ is indeed non-negative because $a \ge b \ge 0$.
Finally, we arrive at
\begin{align}
\mathcal{K} &=  \frac{4}{\sqrt{a-b}} 
\ \sqrt{\frac{a-b}{a+b}}
\  K\qty( \sqrt{\frac{2b}{a+b}} ) 
\nonumber\\
&= \frac{4}{\sqrt{(s+s')^2 + (z-z')^2}} 
\ K\qty( \sqrt{\frac{4 s s'}{(s+s')^2 + (z-z')^2}} ), \nonumber
\end{align}
where $K$ is the complete elliptic integral of the first kind. The SN equation now simplifies to two dimensions as
\begin{align}
i \hbar & \pdv{t} \psi(s,z,t) = 
 - \frac{\hbar^2}{2 m} \nabla^2 \psi(s,z,t) 
\nonumber\\
& -  Gm^2 \int s' ds'dz' 
\Big( |\psi(s',z', t)|^2 +  |\psi(s',-z'-L, t)|^2 \Big) \, \mathcal{K}(s+s',|z-z'|)
\, \psi(s,z,t) . \nonumber
\end{align}

\section{Average kinetic energy}
\label{appendix:AverageKE}

The change of kinetic energy can be calculated from Ehrenfest's theorem:
\begin{align}
\frac{d}{dt}\langle T\rangle & =-\frac{i}{\hbar}\langle[T,H]\rangle
\nonumber \\
 & =-\frac{i}{\hbar}\Big(\frac{-\hbar^{2}}{2m}\Big)\langle[\nabla^{2},V]\rangle
 \nonumber \\
 & =\frac{i\hbar}{2m}\int d\vec{r} \ \psi^{*}(\vec{r}) \ (\nabla^{2}V-V\nabla^{2}) \ \psi(\vec{r})
 \nonumber \\
 & =\frac{i\hbar}{2m}\int d\vec{r} \ \psi^{*}(\vec{r}) \ \Big( \nabla^{2}\big(V\psi(\vec{r})\big) -V\nabla^{2}\psi(\vec{r}) \Big) \ 
 \nonumber \\
& =\frac{i\hbar}{2m}\int d\vec{r} \ \psi^{*}(\vec{r}) \ \Big( \nabla^{2}V \ \psi(\vec{r}) + V \ \nabla^{2}\psi(\vec{r}) + 2\nabla V \cdot \nabla\psi(\vec{r}) - V\nabla^{2}\psi(\vec{r}) \Big)
 \nonumber \\
 & =\frac{i\hbar}{2m}\bigg(\int d\vec{r} \ \psi^{*}(\vec{r}) \ \nabla^{2}V \ \psi(\vec{r}) + 2\int d\vec{r} \ \psi^{*}(\vec{r}) \ \nabla V \cdot \nabla\psi(\vec{r}) \bigg).
\label{KE1}
\end{align}
The second term can be evaluated through integration by parts:
\begin{align}
\int d\vec{r} \ \psi^{*}(\vec{r}) \ \nabla V \cdot \nabla\psi(\vec{r}) &= \int d\vec{r} \ \nabla \cdot \big[\psi^{*}(\vec{r}) \ \nabla V \ \psi(\vec{r}) \big] - \int d\vec{r} \ \nabla\big(\psi^{*}(\vec{r}) \ \nabla V \big)\psi(\vec{r})
\nonumber\\
 & =-\int d\vec{r} \ \Big( \nabla\psi^{*}(\vec{r}) \cdot \nabla V + \psi^{*}(\vec{r}) \ \nabla^{2}V \Big) \psi(\vec{r}) ,
 \nonumber\\
 & =-\int d\vec{r} \ \nabla\psi^{*}(\vec{r}) \cdot \nabla V \ \psi(\vec{r}) - \int d\vec{r} \ \psi^{*}(\vec{r}) \ \nabla^{2}V \ \psi(\vec{r}),
\end{align}
where we used the Gauss theorem and the square integrability of the wave function to cancel the first integral on the right-hand side of the first line. Inserting this into Eq.~\eqref{KE1}
gives an equivalent expression for $\frac{d}{dt}\langle T\rangle $:
\begin{equation}
\frac{d}{dt}\langle T\rangle 
= -\frac{i\hbar}{2m}\bigg(\int d\vec{r} \ \psi^{*}(\vec
r) \ \nabla^{2}V \ \psi(\vec
r) + 2\int d\vec{r} \ \nabla\psi^{*}(\vec
r) \  \nabla V \ \psi(\vec
r) \bigg).
\label{KE2}
\end{equation}
Adding the two expressions in Eqs.~\eqref{KE1} and~\eqref{KE2} we get:
\begin{align}
\frac{d}{dt}\langle T\rangle & =\frac{i\hbar}{2m}\int d\vec{r} \ \Big(\psi^{*}(\vec{r}) \ \nabla V \cdot \nabla\psi(\vec{r}) -\nabla\psi^{*}(\vec{r}) \cdot \nabla V \ \psi(\vec{r}) \Big)
\nonumber \\
 & =\frac{i\hbar}{2m}\int d\vec{r} \ \Big( \psi^{*}(\vec{r}) \ \nabla\psi(\vec{r}) -\nabla\psi^{*}(\vec{r}) \ \psi(\vec{r}) \Big) \cdot \nabla V
 \nonumber \\
 & = -\int d\vec{r} \ \vec{j}(\vec{r}) \cdot \nabla V,
\hspace{.5cm} 
: \qty{  \vec{j}(\vec{r})=\frac{\hbar}{2mi}\Big(\psi^{*}(\vec{r}) \ \nabla\psi(\vec{r})-\nabla\psi^{*}(\vec{r}) \ \psi(\vec{r}) \Big) 
  },
 \nonumber \\
 & = - \oint V \vec j \cdot \vec da + 
 \int d\vec{r} \ \nabla \! \cdot \! \vec{j}(\vec{r}) \ V ,
 \label{EQ_PARTS_J} \\
 & =
 - \int d\vec{r} \ \dv{|\psi(\vec{r})|^2}{t} \ V,
 \hspace{1cm}
 :\qty{ \nabla \! \cdot \! \vec{j} + \dv{|\psi(\vec{r})|^2}{t} = 0 },
\label{eq_Change_in_KE}
\end{align}
where in~(\ref{EQ_PARTS_J}) we used integration by parts and canceled the first integral because $\vec j$ is proportional to the square-integrable wave function.
For the last line we note that the continuity equation holds also in SN model.

Since the time evolution of the probability density in (\ref{eq_Change_in_KE}) is smooth, the finiteness of $\frac{d}{dt}\langle T\rangle$ is determined by the finiteness of the potential $V$. For Gaussian wave functions we showed in~\ref{appendix:SN-Pot-InitialTime} that the potential scales as $\mathrm{erf}(\vec r - \langle \vec{r} \rangle) / |\vec r - \langle \vec{r} \rangle|$ and the finiteness is seen from the Taylor expansion of the erf function (the first term already cancels the denominator).
One can go further to any (physically reasonable) smeared wave functions (not point-like distributions) as follows. Consider first those vectors $\vec{r} \,'$ in the integration in the potential that are far away from vector $\vec r$, say $| \vec r - \vec{r} \,'| > r_0$. Here we expect the potential to be finite and it is easy to see that 
\begin{equation}
\int_{>} d \vec{r} \,' \frac{|\psi(\vec{r} \,')|^2 + |\psi(- \vec{r} \,' - \vec L)|^2}{|\vec r - \vec{r} \,'|} < \frac{1}{r_0} \int_> d \vec{r} \,' \left( |\psi(\vec{r} \,')|^2 + |\psi(- \vec{r} \,' - \vec L)|^2 \right) \le \frac{2}{r_0}
\end{equation}
is bounded for any $r_0$ larger than zero.
Consider now the opposite regime, $| \vec r - \vec{r} \,'| \le r_0$, i.e. vectors $\vec{r} \,'$ within a sphere of radius $r_0$ centered at $\vec r$.
For any bounded probability density functions $|\psi(\vec{r} \,')|^2 + |\psi(- \vec{r} \,' - \vec L)|^2 \le 2 |\psi|^2_{\max} < \infty$ (this is where point particles are excluded) we have
\begin{equation}
\int_{\le} d \vec{r} \,' \frac{|\psi(\vec{r} \,')|^2 + |\psi(- \vec{r} \,' - \vec L)|^2}{|\vec r - \vec{r} \,'|} \le 2 |\psi|^2_{\max} \int_{\le} \frac{d \vec{r} \,'}{|\vec r - \vec{r} \,'|} = 4 \pi r_0^2 |\psi|^2_{\max},
\end{equation}
where the last integral is evaluated in spherical coordinates centered at $\vec r$.
The result is finite for any finite $r_0$ demonstrating that the potential is bounded.

\section{Bipartite covariance matrix}
\label{appendix:CovarianceMatrix}

The covariance matrix formalism is based on the first two statistical moments. Given a bipartite system $AB$ with the canonical operators $\hat u = ( \hat x_A, \hat p_A, \hat x_B, \hat p_B)^T$, the covariance matrix is defined as~\cite{Vidal2002,Adesso2004,Adesso2005}:
\begin{equation}
\bm{\sigma}_{jk} = \bm{\sigma}_{kj}  = \frac{1}{2} \ev{ \hat u_j \hat u_k + \hat u_k \hat u_j }  -  \ev{ \hat u_j }\ev{ \hat u_k } ,
\end{equation}
In the block form it can be rewritten as:
\begin{equation}
\bm{\sigma}  =
\mqty(
\bm{\alpha} & \bm{\gamma} \\
\bm{\gamma}^T & \bm{\beta}
) ,
\end{equation}
where $\bm{\alpha}(\bm{\beta})$ contains the local mode correlation for $A(B)$, and $\bm{\gamma}$ describes the intermodal correlation. 
In our setting the two particles have the same mass and are prepared in identical initial states, which implies that the local modes are identical at all times: $\bm{\alpha} = \bm{\beta}$.

For central interactions truncated at the quadratic term the exact analytical solutions to the covariance matrix, when starting from the ground state of the harmonic trap of frequency $\omega_0$, are given by~\cite{Ankit2023-QuGrav}:
\begin{align}
\bm{\sigma}_{00}  =& \frac{\hbar}{4m\omega_0} \qty[ 2+\omega_0^2t^2+\qty( 1+\frac{\omega_0^2}{\omega^2} )\sinh^2(\omega t) ],   
\nonumber \\[0.5em]
\bm{\sigma}_{02}  =&  \frac{\hbar}{4m\omega_0} \qty[ \omega_0^2t^2-\qty(1+\frac{\omega_0^2}{\omega^2})\sinh^2(\omega t) ],
\nonumber \\[0.5em]
\bm{\sigma}_{11}   =&  \frac{m\hbar\omega_0}{4} \qty[ 2+\qty( 1+\frac{\omega^2}{\omega_0^2} )\sinh^2(\omega t) ], 
\nonumber \\[0.5em]
\bm{\sigma}_{13} =& - \frac{m\hbar\omega_0}{4} \qty( 1+\frac{\omega^2}{\omega_0^2} ) \sinh^2(\omega t), 
\nonumber \\[0.5em]
\bm{\sigma}_{01} =& \frac{\hbar}{8} \qty[  2\omega_0 t + \qty( \frac{\omega_0}{\omega}+\frac{\omega}{\omega_0} )\sinh(2\omega t) ],   
\nonumber \\[0.5em]
\bm{\sigma}_{03} =& \frac{\hbar}{8} \qty[  2\omega_0 t - \qty( \frac{\omega_0}{\omega}+\frac{\omega}{\omega_0} )\sinh(2\omega t) ],
\end{align}
where $\omega$ encodes the mutual interaction between the two particles as follows.
For an arbitrary central potential expanded in a series in terms of the displacement-to-separation ratio, the parameter $\omega$ is found by equating the coefficient of the quadratic term with $-m\omega^2/4$.
For the Newtonian gravitational potential one finds $\omega^2 = 4Gm/L^3$.

\section{Correlation quantifiers}
\label{appendix:QuantumCorrelations}

Mutual information, which quantifies the departure of a quantum state from a product form~\cite{Nielsen2012}, can be computed from the covariance matrix as~\cite{Serafini2003}:
\begin{equation}
I(\bm{\sigma}) = f\qty(\frac{1}{\hbar}\sqrt{\abs{\bm\alpha}}) + f\qty(\frac{1}{\hbar}\sqrt{\abs{\bm\beta}}) - f\qty(\frac{1}{\hbar}\nu_+(\bm{\sigma})) - f\qty(\frac{1}{\hbar}\nu_-(\bm{\sigma})),    
\end{equation}
with
\begin{equation}
f(x) = \qty(x\!+\frac{1}{2})\log_2\qty(x\!+\frac{1}{2}) - \qty(x\!-\frac{1}{2})\log_2\qty(x\!-\frac{1}{2}) ,
\end{equation}
where $\nu_{\pm}$ are the symplectic eigenvalues of the covariance matrix $\bm{\sigma}$ given by
\begin{equation}
{\nu}_{\pm}(\bm{\sigma}) =  \frac{1}{\sqrt{2}} \sqrt{
{\Sigma}(\bm{\sigma}) \pm \sqrt{{\Sigma}^2(\bm{\sigma}) - 4 \abs{ \bm{\sigma} } } }  ,
\end{equation}
where $\abs{\bm{x}}$ is the determinant of matrix $\bm{x}$ and $\Sigma(\bm{\sigma}) = \abs{\bm{\alpha}} + \abs{\bm{\beta}} + 2 \abs{\bm{\gamma}} $.


The negativity of the partially transposed density matrix is a necessary and sufficient condition for entanglement in two--mode Gaussian states~\cite{Simon2000}. As a result of partial transposition, the covariance matrix is transformed to $\tilde{\bm{\sigma}}$, which differs from $\bm{\sigma}$ by a sign-flip of $\abs{\bm{\gamma}}$~\cite{Adesso2005}. 
The symplectic eigenvalues of the partially transposed covariance matrix are given by~\cite{Vidal2002,Adesso2004}:
\begin{equation}
\tilde{\nu}_{\pm}(\bm{\sigma}) =  \frac{1}{\sqrt{2}} \sqrt{
\tilde{\Sigma}(\bm{\sigma}) \pm \sqrt{\tilde{\Sigma}^2(\bm{\sigma}) - 4 \abs{\bm{\sigma}} } } ,
\label{EQ_SYMPARTIAL}
\end{equation}
where $\tilde{\Sigma}(\bm{\sigma}) = \abs{\bm{\alpha}} + \abs{\bm{\beta}} - 2 \abs{\bm{\gamma}} $. Entanglement is quantified by the minimum symplectic eigenvalue via logarithmic negativity:
\begin{equation}
E(\bm{\sigma}) = \max \Bigg[ 0, \ -\log_2\qty(  \frac{\tilde\nu_-(\bm{\sigma})}{\hbar/2} ) \Bigg].
\label{eq:E_from_covmat}
\end{equation}


\providecommand{\newblock}{}

\end{document}